 \documentclass[doublecol]{epl2} 
\usepackage{xcolor}
\usepackage{amsmath} 
\usepackage{amsfonts} 
\usepackage[normalem]{ulem}

\newcommand{\ket}[1]{|#1\rangle}
\newcommand{\ketbra}[2]{|#1\rangle\!\langle#2|}
\usepackage{hyperref} 
\begin{document}
\title{Continuous-time quantum walks on dynamical percolation graphs}
\shorttitle{Continuous-time quantum walks on percolation graphs} 
\author{Claudia Benedetti\inst{1}\thanks{E-mail: \email{claudia.benedetti@unimi.it}} \and Matteo A. C. Rossi\inst{2}\thanks{E-mail: \email{matteo.rossi@utu.fi} }
\and Matteo G. A. Paris\inst{1}\thanks{E-mail: \email{matteo.paris@fisica.unimi.it}}}
\shortauthor{C. Benedetti \etal}
\institute{\inst{1}Quantum Technology Lab, Dipartimento di Fisica 
'Aldo Pontremoli', Universit\`a degli Studi di Milano, I-20133 Milano, 
Italy \\ 
\inst{2}QTF Centre of Excellence, Turku Centre for Quantum Physics,
Department of Physics and Astronomy, University of Turku , FI-20014 
Turun Yliopisto, Finland}
\abstract{
We address {\em continuous-time} quantum walks on graphs in the 
presence of   time- and space-dependent noise. Noise is modeled 
as generalized dynamical percolation, i.e. classical time-dependent 
fluctuations affecting the tunneling amplitudes of the walker. In 
order to illustrate the general features of the model, we review 
recent results on two paradigmatic examples: the dynamics of 
quantum walks on the line and the effects of noise on the 
performances of quantum spatial search on the complete and 
the star graph. We also discuss future perspectives, including 
extension to many-particle quantum walk, to noise model for 
on-site energies and to the analysis of different noise 
spectra. Finally, we address the use of quantum walks as a 
quantum probe to characterize defects and perturbations 
occurring in complex, classical and quantum, networks.}
\maketitle
\section{Introduction}
Quantum walks (QWs) describe the propagation of a quantum particle
over a discrete set of positions. QWs are the quantum counterpart 
of the classical random walks, i.e. systems where a walker moves on 
a lattice by hopping through sites according 
to a certain set of transition probabilities. A well known example 
is provided by the random walk on the line, where at each time step 
the walker moves according to the tossing of a coin, e.g., it moves 
to the left if the outcome is head and to the right if it is tail.  
In the quantum analogue of the random walk, the evolution is governed by 
a {\em quantum coin}, which may exist in a superposition of
head and tail states, making the propagation of the walker {\em coherent}, 
i.e. evolving in a superposition of possible positions.
The dynamics is discrete in time, each temporal step corresponding to 
a toss of the quantum coin. For this reason this model is named 
{\em discrete-time quantum walks} (DTQW) \cite{aharonov93}.
A different model has been suggested few years later \cite{fahri98},
in which the walker moves continuously in time, in a closer analogy 
with the evolution of classical Markov chains. This model, in 
which the evolution of the walker is governed by a lattice 
Hamiltonian, is usually referred to as {\em continuous-time 
quantum walk} (CTQW).
\par
The concept of QW is naturally connected to the
notion of graph. Indeed a QW, both discrete- and continuous-time,
evolves on a discrete position space, where the states can be 
identified with the nodes of a graph. The edges of the graph 
are then associated with the tunneling amplitudes
between connected nodes. Different graph topologies 
then lead to different dynamics for the walker.
QWs were proven useful tools for several tasks, ranging from 
universal quantum computation \cite{childs09}, transport 
on networks \cite{blumen11,tamascelli16}, quantum algorithms
\cite{gold04,kendon06,farhi08,gamble10,cottrell14}, quantum modelling of 
biological systems, \cite{mohseni08,hoyer10}, graph matching 
\cite{pr09}, and as quantum probes for the topology of graphs \cite{seveso18}.
QWs have been experimentally implemented on different platforms, 
e.g. trapped ions \cite{schmitz09,matjeschk12}, nuclear spins  
\cite{du03} and optical systems \cite{perets08,sansoni12,bian17}.
In realistic implementations of QWs, environmental noise and 
defects may affect the behavior of the quantum walker 
\cite{schreiber11}.  As a consequence, the speed-up observed 
in certain computational tasks may be lost, and the QW may either 
transform into a classical random walk, or localize over few sites \cite{romanelli05,yin08,keating07,amir09,jackson12}.
\par
In this paper, we address the most relevant form of perturbation that 
may affect a graph: percolation. In a percolation graph links between 
nodes are created with a certain probability $p$. A generalization of the static
percolation, where the links can be created and destroyed in time with a certain rate, is called dynamical percolation \cite{druger83}. If the percolation 
rate vanishes, the static case is recovered.
For CTQW, the absence or presence of a link between two nodes of 
the underlying graph is identified with the corresponding tunneling 
amplitude between the walker sites, which may take a zero or non-zero 
value, i.e. it can flip between two values.
This duality allows us to further generalize the percolation model
by assuming that the coupling between sites can
randomly switch between any two non-zero values, thus mimicking the 
fact that the weights of the edges are dichotomic random variables. 
We call this dynamical noise \emph{generalized percolation}, since 
it includes dynamical percolation as a special case. In particular, 
a convenient way to describe generalized percolation is by means
of the random telegraph noise (RTN): a stochastic process where
a certain variable may flip between two values at a certain rate, 
that from now on we refer to as {\em percolation rate}.
\par
The aim of this perspective article is to describe realistic models 
of quantum walks affected by of noise. In particular, we focus 
on the CTQW model in the presence of generalized percolation described 
by RTN. The aim is to provide a general understanding of the 
role of environmental noise in the dynamics of the walker by 
reviewing recent results concerning the temporal behavior of 
a CTQW, with particular attention to the propagation properties of the walker 
and on its ability to search a marked vertex on a graph. 
The paper is organized as follows: we first establish the notation 
and we introduce concepts of graph Laplacian and CTQW Hamiltonian;
After that, we introduce noise in the model. We then review recent results 
on the propagation properties of the walker in the presence of
noise and on its ability to fast searching for a target node on 
a graph. We close the paper with concluding remarks and future 
perspectives. 
\section{Dynamics of  CTQW on graphs}
CTQWs evolve on graphs, i.e.  set of $N$ nodes (discrete positions)
connected by edges. If two nodes are connected by a link, then the
walker may jump from one to the other, and viceversa, with a tunnelling 
amplitude $J$. The Hilbert space of the walker is thus spanned by the 
orthonormal position states $\{\ket{j}\}_{j=1}^N, $where $\ket{j}$ 
denotes the state of the the walker localized at site $j$.
The mathematical object that fully characterizes the topology of a 
graph is its adjacency matrix, whose elements are $A_{jk}=1$ if 
nodes $j$ and $k$ are connected, and $A_{jk}=0$ otherwise, i.e. 
if there is no edge linking $j$ and $k$. From the adjacency matrix 
it is possible to build the Laplacian $L$ of the graph: $L_{jk}=A_{jk}$ 
if $j \neq k$ and $L_{jk}=-d_k$ if $j=k$, where  $d_j=\sum_{k}A_{jk}$ is 
the so-called {\em vertex degree}.
The Hamiltonian for a CTQW on a graph is thus defined by
\begin{equation}
\label{ham} 
\mathcal{H}=-J_0\, L\,.
\end{equation}
An initial state of the walker $\ket{\psi_0}$ evolves according to
$\ket{\psi_t}=e^{-i \mathcal{H} t} \ket{\psi_0}$, where we set $\hbar=1$.
The evolution through the Laplacian operator $L$ is one possible generator 
for the CTQW dynamics. But since quantum mechanics only imposes that 
the  Hamiltonians are Hermitian operators,  another possible
candidate to describe the evolution of the walker is the  adjacency 
matrix $A$ alone, leading to the Hamiltonian $H'=-J_0 A$.
In the case of regular graphs, where the vertex degrees
are all equal, the two Hamiltonians $H$ and $H'$ only differ 
for a term proportional to the identity matrix, thus they generate  
equivalent time evolutions, while this equivalence does not hold true for
irregular graphs. The different dynamics generated by these Hamiltonians
and the physical systems that they are associated with are thoroughly 
described in Ref. \cite{wong16}. In the following, we will focus on the 
evolution generated by the Laplacian.
\par
In the simple case of   the line, i.e. a one-dimensional
regular graph, the Hamiltonian reads
\begin{equation}
\mathcal{H}_{\text{\tiny L}}=2J_0\sum_j \ketbra{j}{j}-J_0\sum_j\Big(\ketbra{j}{j+1}+\ketbra{j+1}{j}\Big)\,,
\end{equation} 
which physically corresponds to the  propagation of a particle in a periodic 
potential, e.g. to simulate tight-binding models \cite{cuevas11}.
Despite the simplicity of the underlying graph, this model allows us to 
highlight the differences between the quantum and the classical QW. The 
most striking difference is the limit distribution of the particle for 
long times: in the case of a classical walk, the transition probability 
from the site $j$ to the site $k$ may be expressed as $p_{kj} (t) 
=\langle k | e^{- \mathcal{H} t}|j\rangle$ and thus, due to the central
limit theorem, the long-time probability distribution of the walker 
is Gaussian, while for a CTQW
a non-trivial non-Gaussian shape is found \cite{konno05}.
Indeed, the probability of finding the quantum particle at site $k$ at time $t$ 
when it is initially localized at site $k_0=0$ is $p_k(t)=J_{|k|}^2(2J_0 t)$, 
where $J_k(x)$ is the Bessel function of order $k$. This probability 
distribution has many peaks, with the external larger than the internal 
ones, and it is symmetric with respect to the central point $k=0$.
\par
An interesting characteristic of CTQW is that it  spreads on the infinite line with a variance $\sigma_q^2\propto t^2$ (referred to as \emph{ballistic} propagation), while in the classical case the variance is $\sigma_c^2\propto t$ (\emph{diffusive} propagation), meaning that a quantum walker is able to explore the nodes faster that
 the classical one. This property has sparked research into possible applications of QW for computational and transport tasks.
\section{Spatial search}
The ballistic propagation of CTQW has been suggested as a resource 
to improve the search for a marked node on a graph, a task requiring
a time of order $\mathcal{O}(N)$ by classical, diffusive, 
propagation. The corresponding quantum CTQW search algorithm has been 
introduced in \cite{childs09} by means of the Hamiltonian
 \begin{equation}
 \mathcal{H}^{ s}=-J \,L- \ketbra{w}{w}\,,
 \end{equation}
which is expected to drive the walker to the target node $\ket{w}$, 
with the help of the oracle operator $\ketbra{w}{w}$.
The coefficient $J$ is the tunnelling amplitude between any 
two connected nodes, and it needs to be optimized  in order to yield the maximum 
probability of finding the walker on the target node, given 
that the walker is initially prepared in a superposition of all 
sites, i.e. the maximum of 
\begin{equation}\label{eq:psucc}
 p_w(t)=\left|\langle w|e^{-i  \mathcal{H}^{s} t}\ket{s}\right|^2,
\end{equation}
where $\ket{s}=\frac{1}{\sqrt{N}}\sum_{j}\ket{j}$. 
For few special 
regular graphs, it has been proved \cite{childs09} that the algorithm 
finds the target state (i.e. the walker localises on the target) in 
a time of order of $\mathcal{O}(\sqrt{N})$, quadratically faster
than the classical analogue.
For the {\bf complete graph} $C_N$, i.e. a graph where each of the 
$N$ nodes is connected to all the other nodes, it was demonstrated  
that CTQW search is equivalent to the Grover algorithm \cite{farhi98}, 
and yields a unit probability of finding the target after a 
time $T=\frac{\pi}{2}\sqrt{N}$, for any $N$.
The proof of this result is obtained by setting $J=\frac{1}{N}$ and
by working on the reduced two-dimensional subspace spanned by the 
vectors $\{\ket{r},\ket{w}\}$, where $\ket{r}=\frac{1}{\sqrt{N-1}}
\displaystyle\sum_{k\neq w}\ket{k}$.
The reduced  search Hamiltonian for the complete graph can 
thus  be written as:
\begin{equation}
\mathcal{H}_{\textsc{\tiny $C_{\!N}$}}^s=\frac{1}{N}\begin{pmatrix}
1&-\sqrt{N-1}\\
-\sqrt{N-1}&-1
\end{pmatrix},
\label{redhc}
\end{equation}
and the initial state $\ket{s}=\sqrt{\frac{N-1}{N}}
\ket{r}+\sqrt{\frac{1}{N}}\ket{w}$, such that 
$\mathcal{H}^s_{\textsc{\tiny $C_{\!N}$}}\ket{s}=-\frac{1}{\sqrt{N}}\ket{w}$.
Upon exploiting the fact that
$(H^s_{\textsc{\tiny $C_{\!N}$}})^k=\frac{1}{N^{\left\lfloor\! k/2\!
\right \rfloor}}\,(H^s_{\textsc{\tiny $C_{\!N}$}})^{\frac{1-(-1)^k}{2}}$
with $(H^s_{\textsc{\tiny $C_{\!N}$}})^0=\mathbb{I}$,
the probability  of finding the target node is found to be
$p_w(t)=|\langle w|e^{-iHt}|s\rangle|^2=\frac{1}{{N}}\cos^2\left(\frac{t}{\sqrt{N}}\right)+\sin^2\left(\frac{t}{\sqrt{N}}\right)$,
i.e. the algorithm finds $\ket{w}$ with probability one in a 
time $t^{\text{\tiny opt}}=T=\frac{\pi}{2}\sqrt{N}$. 
\par
Recently, the same quadratic speedup has been proved also for the 
{\bf star graph} \cite{cattaneo18}, i.e. a graph where only a 
central node is connected to all the other $(N-1)$ nodes. In this case, 
two different scenarios may be considered: either the target is the 
central node or an external one. In the first case, it can be shown 
that the reduced search Hamiltonian
has the same  form as the one for the complete graph in Eq. 
\eqref{redhc} in the $\{\ket{r},\ket{w}\}$ basis. It follows that, 
despite the completely different topology, the reduced dynamics of 
$p_w(t)$ is the same as in the complete-graph case,
with a maximum equal to one reached in time $t^{\text{\tiny opt}}=T$.
The analogy with the complete graph is broken if the target is an 
external node. In this case the reduced space is made of the three 
states $\{\ket{c},\ket{w},\ket{r}\}$, where $c$ stands for the 
central node and $\ket{r}=\frac{1}{\sqrt{N-2}}\sum_{k\neq c,w}
\ket{k}$. The reduced
Hamiltonian for the star graph with external target and coupling 
$J=1$ reads:
\begin{equation}
\mathcal{H}^s_{\textsc{\tiny star}}=\begin{pmatrix}
N-1&-1&-\sqrt{N-2}\\
-1&0&0\\
-\sqrt{N-2}&0&1
\end{pmatrix}.
\end{equation}
By properly manipulating the expression of the Hamiltonian 
and after using perturbation theory \cite{cattaneo18},
one obtains that the initial state $\ket{s}$ evolves into 
the state $\ket{w}+\mathcal{O}(N^{-1/2})$ after a time 
$t^{\text{\tiny opt}}=T$. This indicates that for very 
large values of $N$ the algorithm is optimal even for external 
target nodes. Moreover, numerical simulations show that the success
probability for a smaller number of nodes is proportional to $p_w(t^{\text{\tiny{opt}}})\simeq 1-N^{-2}$
with $t^{\text{\tiny{opt}}}\propto\sqrt{N}$: the algorithm 
is successful, with high probability, also for smaller 
values of $N$.
\section{Noisy CTQW} 
In order to address how the dynamics of CTQW is modified by 
graph imperfections or by the interaction with the
environment, let us consider a graph made of nodes of a 
physical network, that may be affected by external noise, i.e., 
turbulences, thermal fluctuations, or imperfections in the 
fabrication process. As a consequence, links may be weakened 
or temporary removed from the graph and the values of tunneling
amplitudes between any two nodes may fluctuate in time.
We are interested in how this noise  modifies the features 
of the QW. 
\par
The Hamiltonian describing this non-ideal CTQW reads:
\begin{equation}
\label{hpert}
\tilde{\mathcal{H}}(t)\!=\!\!\!\sum_{j,k=1}^N
\!\!\Big[\!\! \left(J_{\text{\tiny$0$}}d_{\text{\tiny$j$}}\!+\!\nu J^{\textsc{\tiny S}}_{\text{\tiny$j$}}(t)\right)\!\delta_{\text{\tiny$j\!k$}} \!-\!\left(J_{\text{\tiny$0$}}\!+\!\nu J^{\textsc{\tiny T}}_{\text{\tiny$j\!k$}}(t) \right)\!A_{\text{\tiny$j\!k$}}\Big] \ketbra{j}{k}\!\!
\end{equation}
where $J^T_{jk}(t)$ and $J^S_{j}(t)$ are adimensional stochastic processes
that describe the perturbation of the
tunneling and  the on-site energies respectively. The matrix $J^T$ is 
symmetric, whereas $\nu\in [0,J_0]$ ia a real parameter which determines 
the strength of the noise. The factors $A_{jk}$ are the elements of the 
adjacency matrix of the graph and $\delta_{jk}$ is the Kronecker delta.
\par
The Hamiltonian \eqref{hpert} is the most general expression of QW in 
the presence of  classical noise: it contains perturbations on both the 
diagonal and off-diagonal elements.
In general, the coefficients $J^T(t)$ depend on time, and describe 
random fluctuations in the tunnelling amplitudes (dynamical percolation). 
The autocorrelation function of the noise dictates the characteristic 
time of the perturbations $\tau_c$. Two regimes arise: {\it fast} noise 
if $\tau_c<1/\nu$ and {\it slow} noise in the opposite case, $\tau_c>1/\nu$. 
In the limiting case $\tau_c\rightarrow \infty$ we have {\it static} 
noise (ordinary percolation) which is apt to describe defects in the graph, 
e.g. due to impurities or imperfections during the implementation of the
couplings between nodes.
\par
In order to describe dynamical percolation, we should model a situation 
where links are created ad destroyed randomly in time with a certain 
percolation rate \cite{darazs13}. This may be obtained
assuming that the links are affected by random telegraph noise (RTN),
which is a non-Gaussian stochastic process where a 
random variable $X$ switches in time between two values, e.g. $X=\pm 1$, 
with a certain switching or percolation rate $\gamma$. The probability 
that $X$ switches $n$ times in a time $t$ follows a Poissonian distribution 
with mean value $\bar{n}=\gamma t$. The autocorrelation function of the 
noise is exponential $C(t)=e^{-2\gamma t}$, corresponding to a Lorentzian 
spectrum. If the couplings $\{J_{jk}^T(t)\}$ in Eq.\eqref{hpert} are 
independent realizations of RTN with $J_{jk}^T(t)=\pm1$, then the tunneling 
energies  (i.e. the links of the  graph)  jump in time between the values 
$J_0\pm \nu $. If $\nu=J_0$ we recover the true dynamical percolation case, 
where links are created and destroyed with rate $\gamma$. For other values 
of $\nu$ we have generalized  dynamical percolation, in which links, 
rather than just appearing and disappearing in time, are {\em modulated}: 
the coupling constants switch  between  a larger and a smaller non-zero 
value or, in other words, they are weakened and strengthened randomly in time.
\par
The dynamics of the noisy walker is described as an ensemble average over all possible realizations of  $\{J^{\text{\tiny T}}(t)\}$,
\begin{equation}\label{eq:quantum_map}
  \rho(t)=\langle U(t)\rho_0 U^{\dagger}(t)\rangle_{\{J^{ \text{\tiny T}}\}},
\end{equation}
where $U(t)=\mathcal{T} \exp\left[{-i\int_0^t \mathcal{H}(s)ds}\right]$ with $\mathcal{T}$ the time-ordering operator and
$\rho_0$ the initial state of the walker.
Equation \eqref{eq:quantum_map} describes a completely positive, trace-preserving quantum map. The evolved density matrix $\rho(t)$ cannot be, in general, 
computed analytically, and numerical techniques are required.
For a low number of nodes and noise sources, an exact method using a 
quasi-Hamiltonian technique is available \cite{rossi18}, but for large 
number of nodes the ensemble average over the noise realizations has to 
be performed with Montecarlo techniques, possibly using GPUs for efficient 
parallel computation \cite{piccinini17}.
\section{Noisy CTQW dynamics}
\begin{figure}
  \includegraphics[width=0.9\columnwidth]{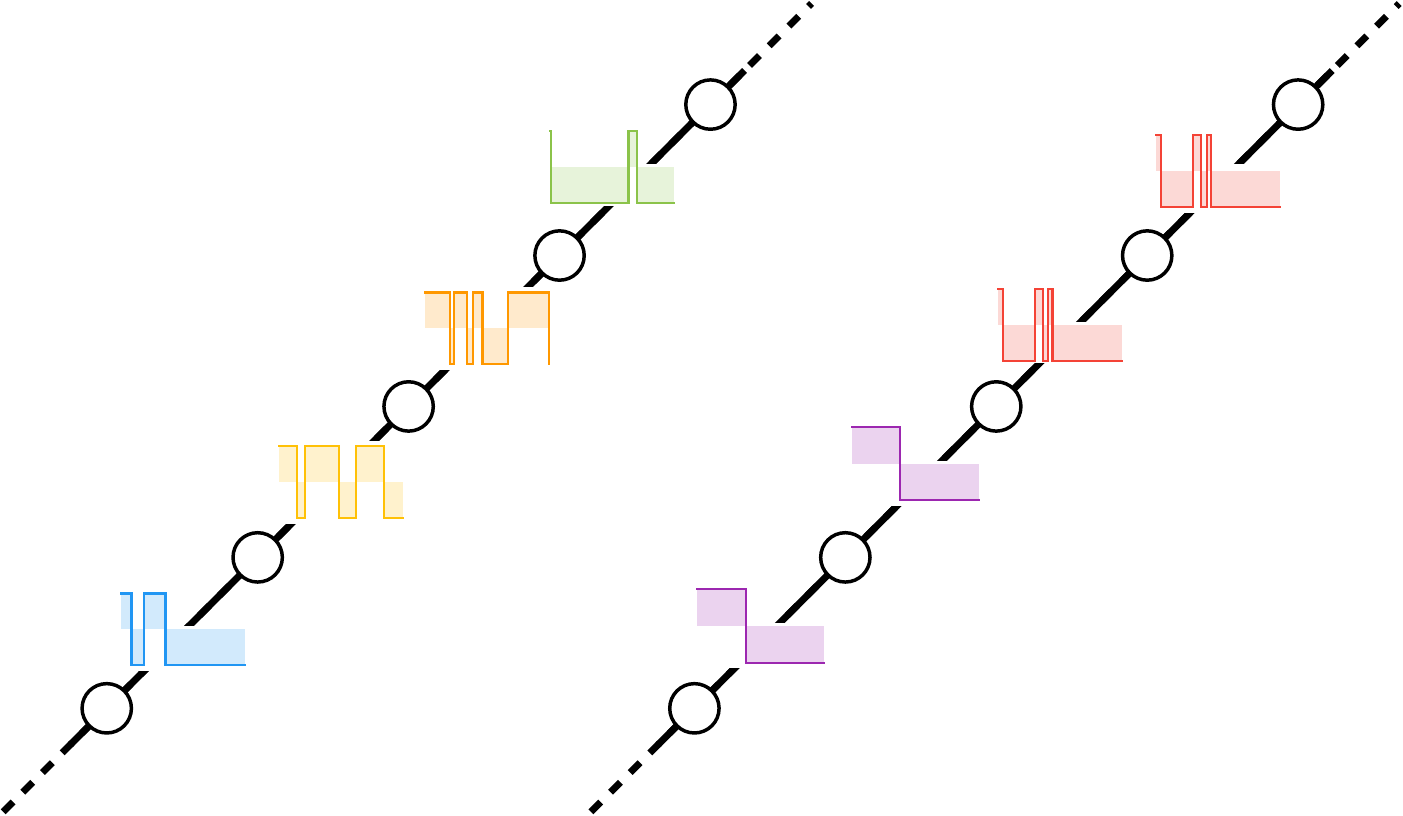}
  \caption{Pictorial representation of the lattice described in Eq. \eqref{linepert}, with uncorrelated noise sources (left) and spatially correlated noise (right)}
  \label{fig:qw_line}
\end{figure}

Let us start by discussing recent results on the effects of classical
noise on the dynamics of a CTQW on a simple one-dimensional graph, i.e. 
a line. At first, we want to understand  how the dynamics of the walker 
is changed if noise is  introduced in the model. To this aim, we 
assume a generalized percolation where the links of the graph switch
between two values, and focus attention on CTQW on a line with periodic 
boundary conditions. The noise is introduced as RTN with strength
$\nu$ to the coupling constants. We also set $J^{\text{\tiny S}}(t)=0$, i.e. we focus  
to the off-diagonal perturbation which describe the phenomenon of
percolation. Upon specializing Eq. \eqref{hpert} to the case of a 
line and setting $J_0=1$, i.e. expressing all quantities in unit 
of $J_0$, we obtain
\begin{equation}
\label{linepert}
\tilde{\mathcal{H}}_{\text{\tiny L}}\!=\!\sum_j\! 2\ketbra{j}{j}-
\!\sum_j\!\left[1\!+\!\nu J^{\textsc{\tiny T}}_{j}(t)\right]\!\!
\big(\ketbra{j}{j+1}+\ketbra{j+1}{j}\big)\,.
\end{equation} 
This model, {depicted in Fig.~\ref{fig:qw_line} (left),} has been studied in \cite{benedetti16}, where the different 
perturbations  $J^{\textsc{\tiny T}}_{j}(t)$ are iid realizations of RTN, 
i.e. $\langle J^{\textsc{\tiny T}}_{j}(t) J^{\textsc{\tiny T}}_{k}(0) \rangle\!=\delta_{jk}e^{-2\gamma t}$, where $\gamma$ is the process percolation rate.

\begin{figure}
  \includegraphics{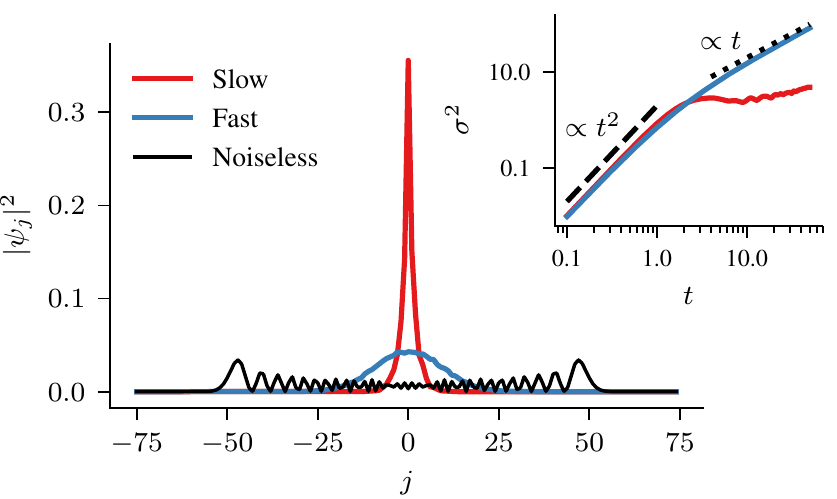}
  \caption{Probability distribution of the walker at $t=50$ for slow ($\gamma = 0.01$, red) and fast ($\gamma= 1$, blue) noise. The noiseless walker is shown in black for comparison. In the inset, the variance $\sigma^2$ as a function of time. The black lines are visual guides for different propagation regimes: ballistic (dashed) and diffusive (dotted). With fast noise we can see a transition from the ballistic to the diffusive propagation, while slow noise causes temporary localization of the walker.}
   \label{fig:qw_line_spread}
\end{figure}

The spread of the particle is analyzed in terms of the variance of the wave function as a function of time.
By increasing the  value of the percolation rate, one is able to move from a localized regime,
where the wave function stays localized over few sites of the chain, to a classical diffusive regime, with a Gaussian-like probability
distribution over the lattice nodes (see Fig. \ref{fig:qw_line_spread}). 
Specifically, in the slow noise regime, also called quasi-static since the  percolation rate is  very small compared to $J_0$,
the larger the strength of the noise $\nu$, the more spatially confined the spatial probability distribution is.
Localization in quantum walks has been largely addressed in the past years. However those models
always considered localization induced by static disorder on the on-site energies of the QW \cite{keating07,amir09,jackson12,croy11} .
Model \eqref{linepert} instead shows that localizations can also be due to quasi-static noise on the
tunneling energies, thus defying the common concept that only disorder can confine a quantum particle.
When the particle localizes, transport through the lattice is suppressed, thus localization is often considered
a threat to transfer of an excitation. However, there are situations where localization
is deliberately induced in order to keep the walker confined into few sites, thus viewing
disorder as a resource more than a threat \cite{sapienza10}.
\par
In the opposite regime of fast noise, a small strength  of the perturbations leads to quasi-unperturbed probability
distribution, while larger values make the walker 'classical', with a Gaussian-shaped distribution, and
the system is driven from a ballistic to a diffusive propagation.
A qualitatively similar behavior is obtained if the walker is initially
prepared in a Gaussian wave packet with a nonzero velocity $p_0$ and standard deviation $\Delta$ i.e.,
\begin{equation}
\ket{\psi_0}=\sum_j\frac{1}{2\pi \Delta^2}e^{-\frac{j-N/2}{2\Delta^2}}e^{-ip_0 j}\ket{j}.
\end{equation}
This indicates
independence of the above results on the initial state of the walker. Moreover, in this case, assuming small strength of
the noise, transport through the lattice is possible.
\par
An advantage of the generalized percolation noise model is that 
it is not specific to the single-particle CTQW. Indeed, it can  easily be
integrated in a $n$-particle QW model, in order to study the effects of disturbance on the many-body dynamics. For the two-particle case, for instance,  
the Hamiltonian is:
\begin{align}
\label{h2p}
&H_{2p}(t)=H_0(t)+H_{\text{\tiny int}}\\
& H_0(t)=\tilde{\mathcal{H}}(t)\otimes\mathbb{I}+\mathbb{I}\otimes\tilde{\mathcal{H}}(t)
\end{align}
where $\tilde{\mathcal{H}}(t)$ is the single particle perturbed Hamiltonian given in Eq. \eqref{hpert}
and $H_{\text{\tiny int}}=H_{\text{\tiny int}}(|j-k|)$ is the interaction Hamiltonian, which  usually 
depends on the distance between particles located at sites $j$ and $k$.
Different dynamical behaviors arise  depending on the statistical nature of the particles, i.e. whether they are
bosons or fermions,  on  their indistinguishability and the noise parameters.  Moreover, the initial conditions and the strength of inter-particle interaction are crucial for their time-evolution.
Generalized percolation  for a two-particle CTQW is
analyzed in \cite{siloi16,siloi17} for on-site and nearest-neighbors interactions. 
Numerical evidence shows that fast percolation leads to a faster propagation 
of the initial wave packet of two interacting particles with respect to the noiseless case
 thus breaking the localization induced by the inter-particle interaction. 
This means that some components of the wave function
 gain a larger momentum because of noise and can travel faster across the lattice 
 introducing a new regime that it is not achievable without noise. 
 This behavior is possible only when the
 particle are initially localized within the range of interaction. In the  slow percolation regime localization is induced, with the particles unable to propagate 
though the lattice.
\par
The model described by Eq.\eqref{linepert} can be further improved by assuming that the
tunneling amplitudes can be grouped into spatial domains, with the constraint that all edges
within the same domain are synchronized in their fluctuations \cite{rossi17}, as depicted in Fig.~\ref{fig:qw_line} (right). These spatial regions are called
percolation domains. In this case spatial correlations are added to temporal correlations 
and the autocorrelation function of the noise becomes
\begin{equation}
C(t)\!=\!
\begin{cases}
e^{-2\gamma t}&\!\!\text{if $j,k$ belong to the same  domain}\\
0 &\text{otherwise}
\end{cases}.
\end{equation}
The spatial domains are created randomly, i.e. if two neighbor edges are correlated with probability $p$,
then the probability of creating $M$ domains follows the distribution
$P_M=\left(\begin{array}{cc} N-1\\M-1\end{array}\right) (1-p)^{M-1}p^{N-M}$.
As a consequence, the average length of the domains
 $\overline{L}=\frac{p^N-1}{p-1}$
moves from the case of independent RTN with $\overline{L}=1$, as described
in Eq.\eqref{linepert}, to the case of uniform noise where all edges percolate synchronously $\overline{L}=N$.
The dynamical evolution of the walker is computed as ensemble average not only on the realizations of the noise,
but also on the realizations of the randomly generated domains $\rho(t)=\langle U(t)\rho_0U^{\dagger}(t)  \rangle_{\{J^{\textsc{\tiny T}}_1\dots J^{\textsc{\tiny T}}_M \}} $.
\par
The dynamics of a Gaussian wavepacket with an initial momentum $p_0$ shows that
the average velocity of the packet decreases with decreasing average lengths and thus the quantum walker
can travel longer across the lattice thanks to spatial correlations. The smaller the value of $\gamma$, the
 faster this decay is, leading to a full localization in the case
$\overline{L}=1$, while the presence of spatial correlations breaks the localization.
For bigger values of $\gamma$, on the other side, the effects of large spatial domains is to allow the
wavepacket to travel across the graph with an almost unaltered form, i.e. the walker is transfered
across the line at fast speed and without losing the information about the initial superposition state. All these results show that spatial correlations can assist the transport of quantum
particles over a linear array of nodes.
\section{Noisy spatial search by CTQW}
\begin{figure}
  \centering
  \includegraphics[width=.85\columnwidth]{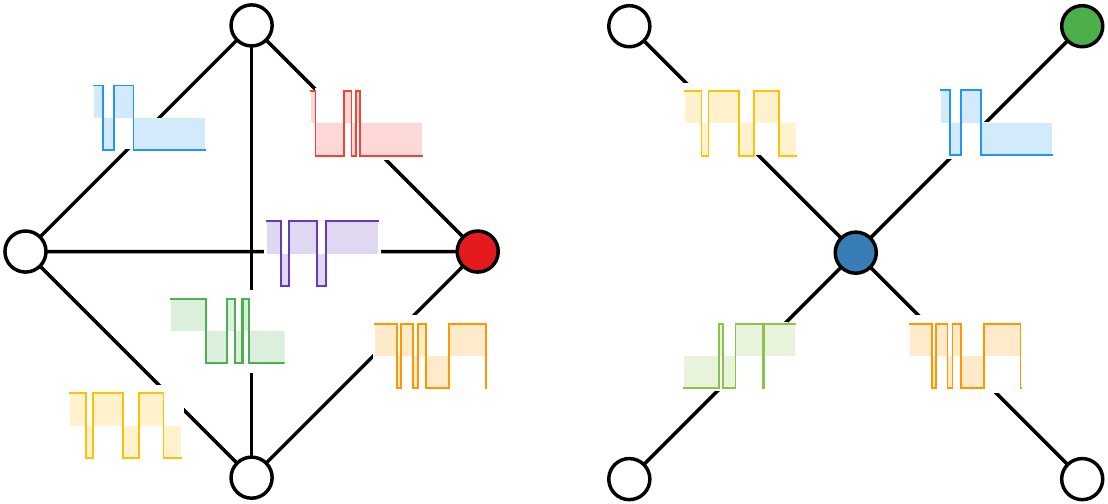}
  \caption{{Pictorial representation of the graphs considered for the 
  quantum spatial algorithm}: the complete graph (left), where each node is connected to all the others, and the star graph (right), with a central node connected to the remaining ones. In the case of the star graph, we have different results for a central target (blue) or an external target (green), as shown in Fig.~\ref{fig:search}.}
  \label{fig:graphs}
\end{figure}

In the following, we report recent advancements on the analysis of the robustness of the spatial search algorithm by CTQW \cite{cattaneo18} against
dynamical percolation by RTN. In order to do so, one needs to compare the success probability, i.e., the maximum of Eq. \eqref{eq:psucc} with respect to time, of finding the target in the noiseless
and noisy case. The study has focussed on the complete graph and on the star graph, 
because they both allow for a quadratic speedup of the search in the noiseless case (as seen above), but have
very different topological properties. A pictorial representation of the model is shown in Fig.~\ref{fig:graphs}. The scaling of the search time with the numbers of nodes $N$ has been studied
for various combinations of noise strength $\nu$ and percolation rate $\gamma$. 
\par
\begin{figure}
  \includegraphics[width=\columnwidth]{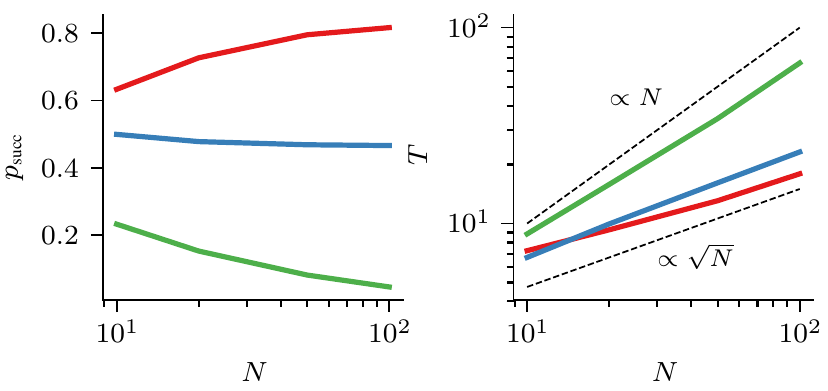}
  \caption{Left: the success probability of the spatial search algorithm as a function of $N$ for slow percolation noise ($\gamma = 0.01$, $\nu = 1.0$) in the complete graph (red), star graph with central target (blue), star graph with external target node (green). On the right, the scaling of the average running time $T$ with $N$ for the same graphs.}
  \label{fig:search} 
\end{figure}

Let us start with the complete graph. Numerical analysis shows that,
depending on the noise regime, different behaviors are found. In particular, for
fast noise, the success probability is very close to one even for percolation noise, while slow noise is detrimental for fast search, with a
decrease in the success probability. This qualitative result do not depend on $N$,
although $p_w(t^{\text{\tiny opt}})$ is slightly higher for larger values of $N$. Moreover, the optimal coupling
$J=1/N$ remains unaltered regardless of the noise. The role of the noise strength is to reduce the success probability,
with full percolation being the worst-case scenario. 
Interestingly, even if the success probability departs from the optimal one, the algorithm still retains an average speedup over the classical one.
Indeed, one can assume that the algorithm can be repeated until the correct result is found, and this happens, on average, after $1/p_\text{\tiny succ}$ times, 
where $p_\text{\tiny succ}$ is the success probability. The average running time of the algorithm is still growing as $\mathcal{O}(\sqrt N)$ even in presence of 
percolation, as shown in Fig. \ref{fig:search}.
\par
Similar results are obtained in the case of the star graph with the target placed on the central node, but with stronger effects of the noise.
Indeed, the optimal coupling is $J=1/N$; the influence of fast noise is almost negligible
while slow noise decreases the success probability. The average running time, however, still scales as $\mathcal{O}(\sqrt N)$, thus the quantum speedup is preserved.
This is not the case if the target node is external. The success probability is in general heavily affected by noise, both in the fast and slow percolation regime, as the left panel of Fig. \ref{fig:search} shows.
Again, the larger the noise strength, the smaller is the success probability. In this case, the average running time,
depending on the strength of the fluctuations, shows a transition from quantum ($\mathcal{O}(\sqrt{N})$) to
classical ($\mathcal{O}(N)$) scaling (see Fig. \ref{fig:search}, right panel).
These results can be interpreted in terms of the connectivity of the two graph topologies: in the complete graph with $N$ nodes there are $N(N-1)/2$ links, 
while in the star graph there are only $N - 1$. While the higher connectivity is not necessary for the noiseless spatial search \cite{meyer15}, it allows for greater redundancy
in presence of noise. In the limiting case of the star graph with external target node, which is connected to a single edge, noise can completely break the algorithm.
\section{Conclusions and perspectives}
The concepts of graph and quantum walk are inherently connected, since a 
CTQW naturally evolves on a graph. In constructing a physical graph, or 
network, defects and noise may come into play, thus deforming and/or damaging 
the original structure. The most simple, yet effective, form of perturbation 
that might affect the topology of a graph is generalized dynamical percolation, 
where the coupling among the nodes fluctuates in time. A special case of this 
noise is ordinary dynamical percolation, in which links are created and removed randomly in time. As a matter of fact, generalized percolation modifies the propagation properties of CTQWs on networks, as well as its performance
in certain quantum information tasks. 
\par
In this perspective article, we have reviewed recent results about the 
effects of generalized percolation on tasks such as transport on a lattice 
and spatial search on graphs. The main results is the observation that 
noise with higher percolation rate leads to faster propagation of the walker.
On the other hand, slow percolation favors localization of the particle. 
This might be a desired behavior in certain situations, but also a drawback 
in others, such as in the spatial search algorithm.
\par
Links are not the only part of a graph that can be affected by noise. 
The on-site energies of the nodes may also experience fluctuations, 
though the corresponding effects are largely unexplored. The presence of 
diagonal defects has been investigated \cite{izaac13}, but a comprehensive 
study of dynamical noise on the on-site energies is still missing. This is 
an interesting topic in itself, since it would allow one to understand the 
role of time-dependent fluctuations compared to static defects, and 
ultimately shed light on the differences and similarities in the dynamics 
induced by diagonal and off-diagonal noise. Besides, RTN, i.e. bistable 
fluctuations, is just one possible model to mimic generalized percolation. 
Indeed, any stochastic process may be employed to effectively describe noise affecting the coupling constants. Relevant examples are the Gaussian version 
of a Lorenzian-spectrum noise, i.e. the so-called Ornstein-Uhlenbeck noise,
and the class of {\em colored} noises, especially the celebrated $1/f$ 
noise, stemming from a weighted collection of bistable fluctuators.
\par
Generalized percolation is a universal noise model, i.e. it is not  
specific to a fixed topology, nor to a single-particle  CTQW.  
Any noisy physical system whose evolution can be mapped into a CTQW on a 
graph \cite{hines} may be described using the noise model discussed in
this perspective article. Moreover, the same stochastic description 
may be applied to 
multi-particle CTQW. Relevant systems where the effects of 
both time  and spatial correlations are worth to investigation 
are those described Hubbard or Fermi models. In those systems, 
besides the study of the role of noise in many-body systems, it
would be of interest to understand the interplay among particle 
interaction, statistics and noise in determining the dynamics 
of the system. As a specific topic of interest, we foresee the 
possible formation of correlated noisy domains, which would 
introduce new features in the multi-particle dynamics that 
are worth exploring.
\par
Another interesting direction for future investigation is the propagation
on hypergraphs, i.e. generalization of graphs where {\em hyperlinks} 
connect two or more nodes, instead of just pairs of nodes as in 
standard graphs. Hypergraphs have been introduced as a more realistic 
description of real networks and, as such, they call for a careful 
noise analysis. So far, studies have been focused on the dynamics of 
discrete-time quantum walks \cite{liu18} and a question arises on whether 
CTQW may be defined on hypergraphs, and how generalized percolation 
affects hyperlinks. Overall, the investigation about the effects of 
disturbance on the dynamics of a CTQW on hypergraphs is a promising
line of research.
\par
As a final remark, we mention that CTQW, which corresponds to a 
quantum particle moving on a random or noisy graph, may be used 
as a quantum probe to characterize the graph and its imperfections. 
In this framework, the added value of using quantum probes
to characterize graphs, and the underlying complex quantum systems, 
is based on the optimisation of the extractable information, as well 
as the inherently small disturbance introduced into the system itself.
In turn, CTQW has been already proved useful to infer the value of 
the coupling constant of a lattice \cite{tamasc16}, and of more complex 
graphs \cite{seveso18}. 
\par
More generally, being able to characterize properties of networks, 
including their noise properties, is an essential step in the 
context of network engineering for quantum information tasks.  
Indeed, searching for imperfections and defects in a physical 
network is a crucial step in the implementation and correct 
functioning of the network itself. In particular, it will be of 
interest in the near future to investigate how to exploit local 
quantum measurements on a controllable quantum probe \cite{nokkala18}
to asses the properties of complex networks instead of resorting 
to global measurements on the whole graph. Understanding whether 
CTQW may be used as reliable probes would imply saving resources, 
such as energy and time, in order to extract precise information 
about large complex networks.
\acknowledgments
The authors are grateful to 
S. Barison, P. Bordone, M. Cattaneo, S. Daniotti,
C. Foti, S. Maniscalco, J. Nokkala, E. Piccinini,
J. Piilo, L. Razzoli, I. Siloi, D. Tamascelli, 
J. Trapani, and P. Verrucchi, for several fruitful 
discussions about quantum walks, complex networks, 
and the effects of classical noise on the dynamics 
of open quantum systems. MGAP is member of INdAM-GNFM.
MACR was supported by the Academy of Finland via the 
Centre of Excellence program (project 312058).

\end{document}